\documentclass[aps,prd,10pt,a4paper,twocolumn,preprintnumbers,nofootinbib,floatfix]{revtex4-1}
\pdfoutput=1
\usepackage[a4paper, hdivide={1.91cm,,1.165cm}, vdivide={1.83cm,,3.4cm}]{geometry}

\usepackage{amstext,amssymb}
\usepackage[intlimits]{amsmath}
\usepackage[ansinew]{inputenc}
\usepackage{graphicx}
\usepackage{units}
\usepackage[hyperfootnotes=false]{hyperref}

\newcommand{\He}[1]{\ensuremath{{}^{#1}\mathrm{He}}}
\newcommand{\aHe}[1]{\ensuremath{{}^{#1}\overline{\mathrm{He}}}}
\newcommand{\anti}[1]{\ensuremath{\overline{\mathrm{#1}}}}

\begin{document}


\preprint{UCI-TR-2019-14}

\title{How to produce antinuclei from dark matter}

\author{Julian Heeck}
\email{Julian.Heeck@uci.edu}
\affiliation{Department of Physics and Astronomy, University of California, Irvine, California 92697-4575, USA}

\author{Arvind Rajaraman}
\email{arajaram@uci.edu}
\affiliation{Department of Physics and Astronomy, University of California, Irvine, California 92697-4575, USA}

\hypersetup{
pdftitle={How to produce antinuclei from dark matter},   
pdfauthor={Julian Heeck, Arvind Rajaraman}
}


\begin{abstract}
We show how to produce antideuteron, antihelium, and other antinuclei in large fractions from the decays of a new particle $\phi$ that carries baryon number. Close to threshold, the production of nuclear bound states is preferred over the decay into individual nucleons, effectively decoupling antinuclei and antiproton fluxes and allowing the former to dominate, in clear contrast to antimatter production via coalescence. $\phi$ can either form dark matter itself or be produced by it, and can give rise to a potentially testable amount of antinuclei.
\end{abstract}

\maketitle


\section{Introduction}

Most of our universe is composed of matter, with only a tiny fraction of antimatter 
produced in highly energetic cosmic events,
measured at experiments such as PAMELA~\cite{Adriani:2008zr}
 and  AMS-02~\cite{Aguilar:2016kjl}.
These experiments also probe Dark Matter (DM) models that can lead to 
an enhanced number of positron and antiproton events~\cite{Silk:1984zy}, and even 
antideuteron~$\anti{d}$ has become a 
promising target~\cite{Donato:1999gy,Brauninger:2009pe,Aramaki:2015pii}.
Future experiments such as GAPS~\cite{Mori:2001dv}
have the potential to  improve these measurements significantly.

Any anomalies in such measurements would be hard to reconcile with known astrophysics and even most physics beyond the 
Standard Model (SM).
This is because  heavier antimatter is generally thought to be impossible to produce in such large numbers from astrophysics or DM, owing to the underlying coalescence models that predict a strong hierarchy of antinuclei fluxes and numbers, very roughly given by~\cite{Poulin:2018wzu}
\begin{align}
N_{\anti{p}} \sim 10^4 N_{\anti{d}} \sim 10^8 N_{\aHe{3}} \sim 10^{12} N_{\aHe{4}} \,.
\label{eq:coalescence}
\end{align}
Antihelium fluxes for example are thus bounded by the 
already observed antiproton flux and the absence of antideuterons~\cite{Carlson:2014ssa,Cirelli:2014qia}.

For instance,  there are  
preliminary observations of AMS-02 of 
six $\aHe{3}$ events and two $\aHe{4}$ events 
in seven years of observations~\cite{ams_talk_2019}.
While these remain to be confirmed, it is already known that it 
is difficult to find theoretical models that
produce such large antihelium fluxes while remaining
consistent with other observations~\cite{Dolgov:1992pu,
Belotsky:2000gx,Mori:2001dv,Bambi:2007cc,Carlson:2014ssa,Cirelli:2014qia,Blum:2017qnn,Coogan:2017pwt,Korsmeier:2017xzj,Poulin:2018wzu,Li:2018dxj,Poulin:2018wzu}.

In this article we will show that there are kinematic regions
where antinucleon production by 
dark matter annihilations and decay can be significant.
In such regions, it is possible to evade the coalescence prediction of Eq.~\eqref{eq:coalescence} by considering \emph{low-energy} processes in which nuclear binding energy and phase space closure become relevant. 
In particular, it allows us to
generate $\anti{d}$ and other 
antinuclei from new physics which may well be related to DM.

We will postulate the existence of  a new particle 
$\phi$ that carries baryon number 
$-k$.
If baryon number is conserved, every decay mode of
$\phi$ must contain $k$ antinucleons.
For a  $\phi$ mass close to the decay
threshold, the outgoing antibaryons will be 
non-relativistic and will have a significant probability of
forming antinuclei with a large mass number $A \leq k$.

We will explore this basic idea and show that we can indeed
build models of new physics where a 
new particle $\phi$ decays with detectable production rates for antinuclei. 
We will focus mostly on producing $\anti{d}$ and
$\anti{He}$ but these models could easily be generalized to the production of
other antiparticles.

The rest of this article is organized as follows: in Sec.~\ref{sec:deuteron} we discuss models that lead to $\anti{d}$ and the possible relation to DM. In Sec.~\ref{sec:helium3} we perform an analogous discussion for the production of $\aHe{3}$ that does not require knowledge of Sec.~\ref{sec:deuteron}. We conclude in Sec.~\ref{sec:conclusions}.
In appendix~\ref{sec:helium4} we provide a discussion of models that produce $\aHe{4}$, which is technically more involved.

\section{Models for \texorpdfstring{$\anti{d}$}{anti-d} production}
\label{sec:deuteron}

As an explicit realization of $\anti{d}$ production
let us introduce a new SM-singlet fermion $\phi$ with baryon number $-2$ and 
lepton number $-1$ (the quantum numbers of the antideuterium atom). 
We assume these 
symmetries to be sufficiently conserved so that they govern the $\phi$ decay modes, with one of the lowest-dimensional operators given by
\begin{align}
\mathcal{L} = \frac{\overline{\phi}^c  {n}\, \overline{p}^c  {e}}{\Lambda^2} 	+ \text{h.c.}
\label{eq:lagrangianD}
\end{align}
Eq.~\eqref{eq:lagrangianD} is expected to be the dominant operator of interest for $\phi$ decay.
We stress that baryon and lepton number are conserved in all interactions underlying Eq.~\eqref{eq:lagrangianD},
so the mediator particles will not induce proton decay, neutron--antineutron oscillations or neutrinoless double beta decay.

\subsection{Decay channels}

We will consider the situation where $m_\phi$ is close to the total
mass of the decay products, i.e.~$m_\phi\sim m_\mathrm{n} + m_\mathrm{p}$, neglecting the electron mass $m_\mathrm{e}$ in the following for simplicity.
In this situation, the nucleons can be treated as point particles and we can ignore
the underlying quark structure. 
The possible final states then depend strongly on $m_\phi$;
if $m_\phi$ is \emph{below} the threshold $m_\mathrm{n}+m_\mathrm{p}$, the 
decay \emph{has} to go into a bound state of the nucleons -- a deuteron with mass $m_\mathrm{d}=m_\mathrm{n}+m_\mathrm{p} - \unit[2.2]{MeV} \simeq \unit[1876]{MeV}$ -- in order to be energetically
 allowed. The only kinematically allowed decay induced by Eq.~\eqref{eq:lagrangianD} for $m_\mathrm{d}< m_\phi < m_\mathrm{n}+m_\mathrm{p}$ is then $\phi \to \anti{d}\,\anti{e}$, whereas heavier $\phi$ have the additional decay channel $\phi \to \anti{n}\,\anti{p}\,\anti{e}$. 
This is then a source for production of $\anti{d}$ as long as the lifetime for
$\phi$ is not too long, which can indeed be the case despite the small available phase 
space, as we will now show.

Assuming $m_\phi$  to be close to $m_{ \mathrm{d}}$ ensures that the nucleons and nuclei 
involved in the process are non-relativistic. 
It is then convenient to calculate the decay rate non-relativistically with 
Fermi's golden rule,
\begin{align}
\Gamma(\phi \to \anti{d} \, \anti{e}) = 2\pi |V(\phi \to \anti{d} \, \anti{e})|^2 \rho \,,
\end{align}
$\rho = (m_\phi-m_\mathrm{d})^2/(2\pi^2)$ being the phase-space density. 
$V(\phi \to \anti{d} \, \anti{e})$ is
the appropriate transition element, here approximated as
 \begin{align}
 \begin{split}
\left|V(\phi \to \anti{d} \, \anti{e})\right|^2 &\simeq \frac{\tfrac12\sum_\text{spins} 
|\mathcal{M}(\phi \to \anti{n}\,\anti{p}\,\anti{e}) |^2}
{2m_\phi \, 2 E_{\mathrm{n}}\,  2 E_{\mathrm{p}}\, 2 E_{\mathrm{e}}}\, |\langle \anti{n}\,\anti{p} | \anti{d} \rangle|^2 \,,
\end{split}
\end{align}
where the first factor is the 
hard process and the second factor is  the overlap of the nucleon wave 
function with the deuteron nucleus.
The  hard process $\phi (k) \to \anti{n}(k_{\mathrm{n}}) \anti{p}
(k_{\mathrm{p}}) \anti{e}
(k_{\mathrm{e}})$
has a  squared matrix element 
\begin{align}
\frac12\sum_\text{spins} |\mathcal{M}(\phi \to \anti{n}\,\anti{p}\,
\anti{e})|^2 = \frac{8}
{\Lambda^{4}}(k_{\mathrm{n}} k  + m_\mathrm{n} m_\phi )(k_{\mathrm{p}} k_{\mathrm{e}}) \,.
\end{align}
To estimate the overlap $\langle \anti{n}\,\anti{p} | \anti{d} \rangle$ of the 
nucleons with the deuteron nucleus wave function we assume that $\phi$ produces the two nucleons initially at a single point, so
 $|\langle \anti{n}\,\anti{p} | \anti{d} \rangle|^2\simeq |\psi (0)|^2$ with the deuteron wave function $\psi (r)$.
In simple shell models, the wave function $\psi (0)$ can be approximately related to 
the radius of d, $|\psi (0)|\sim r_{\mathrm{d}}^{-3/2}$, or simply 
$|\psi (0)|\sim (\unit[100]{MeV})^{3/2}$ as a nuclear-physics energy scale.

The nucleons are non-relativistic but the electrons relativistic, leading to the simple decay rate 
\begin{align}
\Gamma(\phi \to \anti{d} \, \anti{e}) &\simeq \frac{ (m_\phi-m_\mathrm{d})^2}
{\pi \Lambda^{4}} |\psi (0)|^2 \,,
\label{eq:decay_to_d}
\end{align}  
which we expect to be valid for $m_\phi-m_\mathrm{d}< m_\mathrm{d}$.
With $|\psi (0)|^2\simeq (\unit[100]{MeV})^{3}$ this yields the lifetime 
\begin{align}
\tau (\phi \to \anti{d} \, \anti{e}) &\simeq \unit[7]{Gyr} \left(\frac{\unit{MeV}}{m_\phi-m_\mathrm{d}}\right)^2
\left(\frac{\Lambda}{\unit[10^5]{TeV}}\right)^{4} ,
\end{align}
which can easily surpass the age of the universe, 
$t_\text{Universe}\simeq \unit[14]{Gyr}$.

\begin{figure}[t]
\includegraphics[width=0.48\textwidth]{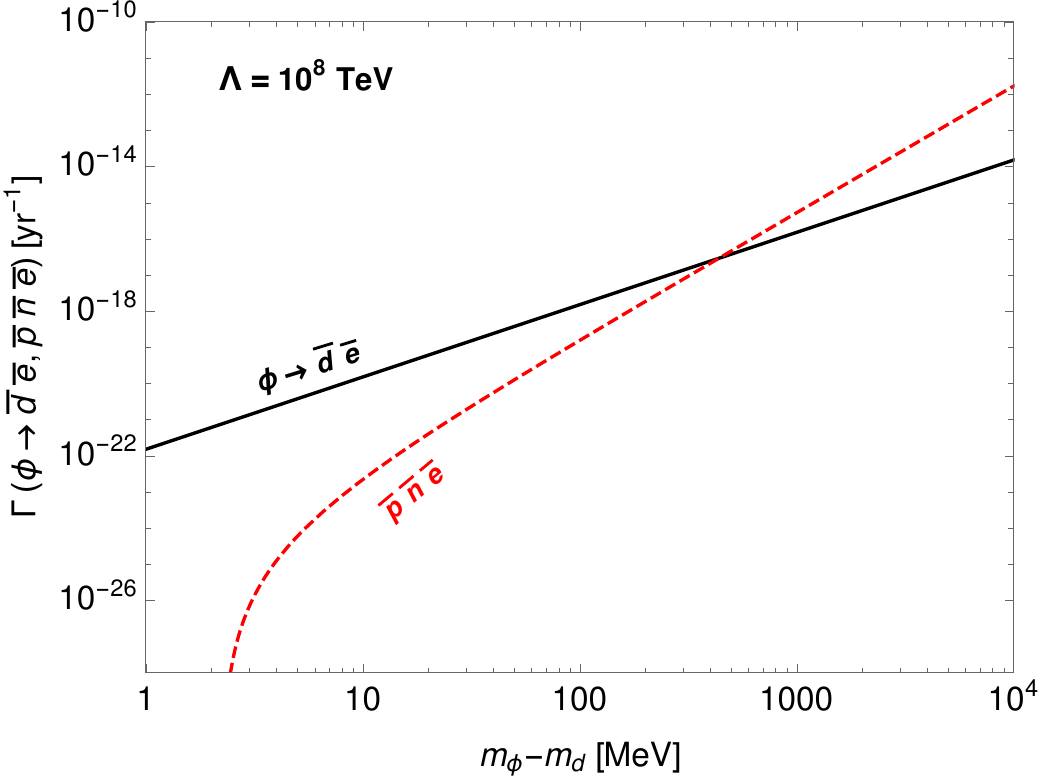}
\caption{Decay rates of $\phi$ to $\anti{d} \,\anti{e}$ and $\anti{p} \,\anti{n} \,\anti{e}$ for $\Lambda=\unit[10^8]{TeV}$.
}
\label{fig:lifetimeD}
\end{figure}

When $m_\phi$ is above the threshold for decay into free nucleons, 
$m_\phi > m_\mathrm{n}+m_\mathrm{p}$, but still small enough to not resolve 
the underlying quark structure or heavier baryons such as $\Sigma$ and $\Delta$, 
$\phi$ will decay into free nucleons with rate
\begin{align}
\Gamma(\phi \to \anti{p} \, \anti{n} \, \anti{e}) &\simeq \frac{\sqrt{2} (m_\mathrm{n}+m_\mathrm{p})^{3/2}\left(m_\phi - m_\mathrm{n}-m_\mathrm{p}\right)^{7/2}}{105\pi^3 \Lambda^4}  .
\label{eq:decay_to_pne}
\end{align}  
This channel starts to dominate over the two-body decay for $m_\phi\gtrsim \unit[2.3]{GeV}$ (Fig.~\ref{fig:lifetimeD})  and  will eventually 
lead to constraints from antiproton-flux measurements that plague most DM models that aim to produce heavier antinuclei~\cite{Carlson:2014ssa,Cirelli:2014qia,Coogan:2017pwt}.

The main conclusion from this analysis is that $\phi$ decays can produce 
almost exclusively antideuteron for $0\lesssim m_\phi-m_{\mathrm{d}} \ll \unit[500]{MeV}$ (see Fig.~\ref{fig:lifetimeD}), while for 
$m_\phi-m_{\mathrm{d}} \sim \unit[500]{MeV}$ we expect 
similar amounts of $\anti{d}$ and $\anti{p}$. For even 
larger $\phi$ masses the $\anti{d}$ fraction will decrease and the 
amount of $\anti{p}$ will become significant, eventually leading back 
to coalescence fractions for the antinuclei.
In order to \emph{dominantly} produce $\anti{d}$ we therefore need $m_\phi \sim \unit[2]{GeV}$, although we could easily consider larger masses before reaching the coalescence fractions of Eq.~\eqref{eq:coalescence}.

\subsection{\texorpdfstring{$\phi$}{phi} as dark matter}
\label{sec:deuteronphiDM}

Since the total lifetime of the new particle $\phi$ can be  longer than the 
age of the universe, it is conceivable that $\phi$ is DM or a subcomponent of it. 
The appropriate relic density
could be produced 
through the underlying mediator interactions that lead to Eq.~\eqref{eq:lagrangianD}. Alternatively, as a result of the assigned baryon number it is reasonable 
to expect $\phi$ to have an \emph{asymmetry} similar to the baryon asymmetry, which 
is consistent with the $\sim\unit[2]{GeV}$ DM mass~\cite{Zurek:2013wia}.
 
Assuming $\phi$ to be DM, we can in principle calculate the antinuclei flux.
As a conservative lower limit on the lifetime we take $\tau > \unit[10^{25}]{s}$~\cite{Slatyer:2016qyl} in order to suppress energy injection during CMB formation from the fast positron that is emitted in $\phi$ decays. For $m_\phi =\unit[2]{GeV}$ this corresponds to a lower limit $\Lambda >\unit[9\times 10^7]{TeV}$.
 There are about 
$10^{12}M_\odot/(\unit[2]{GeV})\simeq 6\times 10^{68}$ DM particles in our galaxy, a fraction
$\unit{yr}/\tau < 3 \times 10^{-18}$ of which decay into $\anti{d}$ per year.
Even with such a long lifetime we can thus produce up to $10^{51}$ $\anti{d}$ nuclei per year in this scenario.
Pushing $\Lambda \to \unit[ 10^{19}]{TeV}$ still yields $\mathcal{O}(10^7)$ antideuteron nuclei per year within the Milky Way, with negligible numbers of antiprotons.
Producing a significant number of antideuterons is thus not difficult in our model.

However, all of these antideuterons are by construction non-relativistic, with 
 kinetic energy $(m_\phi-m_\mathrm{d})^2/(2 m_\mathrm{d})= \unit[4]{MeV}$, far below the 
kinetic-energy threshold for AMS detection of around GeV/nucleon. 
Still, a small fraction of these antinuclei might be accelerated 
by astrophysical processes such as supernovae shock waves, leading to a relativistic 
flux of antideuterons that could be detected by AMS. Most of the remaining 
non-relativistic antinuclei will eventually come in contact with normal 
matter and annihilate, giving rise to characteristic photon spectra that could be detected 
by the Fermi-LAT telescope. A  determination of this fraction is necessary to evaluate whether our model is better constrained by AMS or Fermi-LAT.
Note that we ignored induced-decay processes such as $\mathrm{e}\, \phi\to  \anti{d}\,\gamma$ that could increase the total number of $\anti{d}$ and even boost them, but require a dedicated astrophysical simulation.

\subsection{\texorpdfstring{$\phi$}{phi} produced by dark matter}
\label{sec:deuteronsubDM}

In order to not rely on an astrophysical $\anti{d}$ acceleration mechanism we can imagine $\phi$ to be \emph{produced boosted} by a heavier DM particle $\chi$, either via annihilations or decays (e.g.~$\chi \to \phi\phi$ if 
$\chi$ carries baryon number $-4$ or $\chi \to \phi\phi\phi$ if $\chi$ has baryon number $-6$).
In this basic setup there are no CMB constraints on the lifetimes of $\chi$ or $\phi$, other than $\tau_\chi>t_\text{Universe}$ by assumption.
In particular, the decay rate of $\phi$ is only constrained by the antideuteron flux and UV-considerations for $\Lambda$, which we address later.

\begin{figure}[t]
\includegraphics[width=0.48\textwidth]{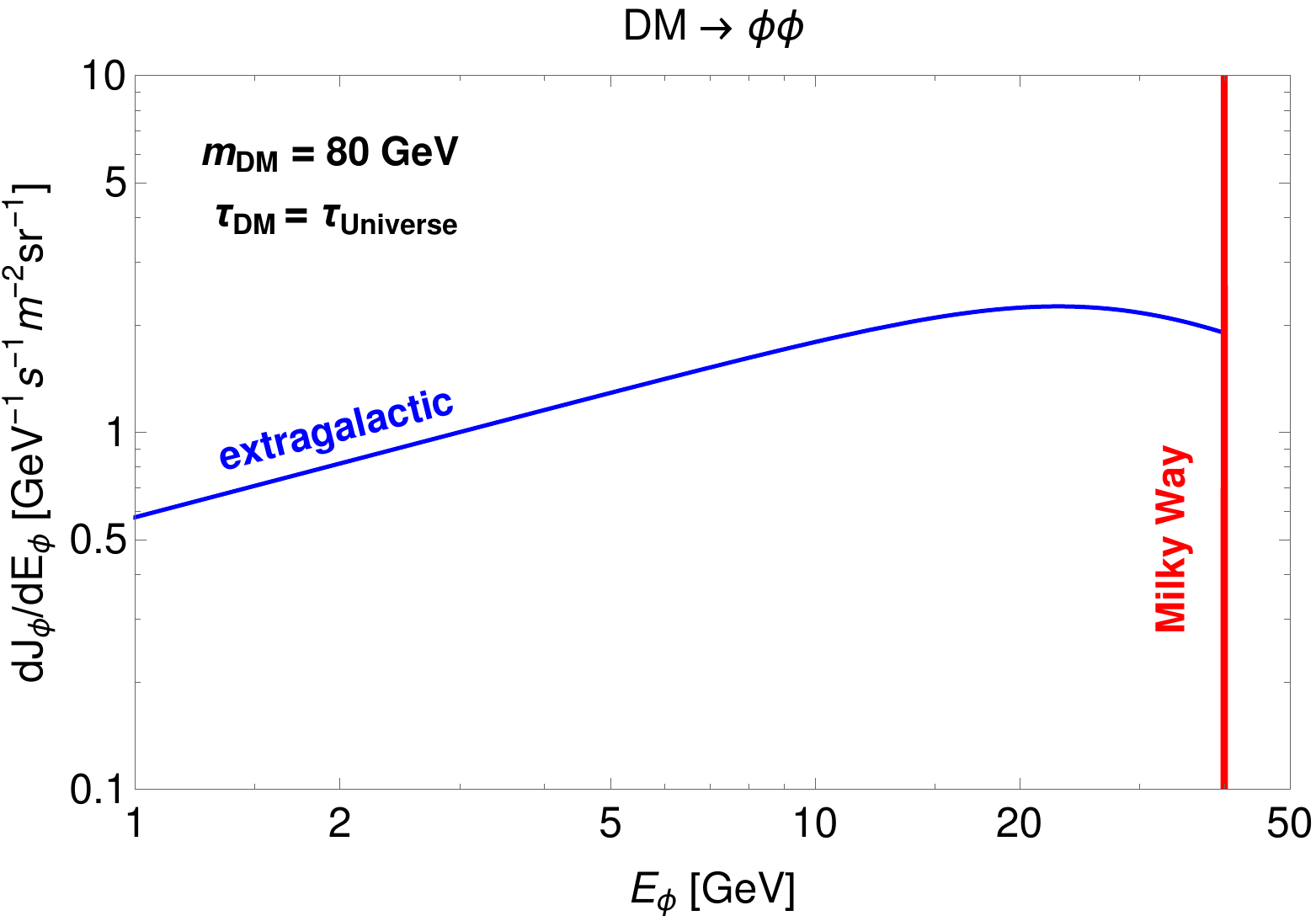}
\caption{Differential flux of $\phi$ as produced by DM decay DM$\to\phi\phi$ of galactic (red) and extragalactic origin (blue).
}
\label{fig:phiflux}
\end{figure}

Consider for example $\chi\to\phi\phi$ in our galaxy with $m_\chi =\unit[80]{GeV}$ and $m_\phi =\unit[2]{GeV}$, which gives a flux of $\phi$ particles with $E_\phi=\unit[40]{GeV}$  at Earth of
\begin{align}
J_\phi &= \frac{2}{4\pi m_\chi \tau_\chi}\int_0^\infty \text{d} s \,\rho_\text{Halo}[r(s)]\sim\frac{10^6}{\unit{m^2\, s\, sr}} \,\frac{t_\text{Universe}}{\tau_\chi} \,,
\label{eq:J_phi_galaxy}
\end{align}
where the integral is over the line of sight~\cite{Esmaili:2012us}. 
Only a small fraction $d/(\tau\, |\mathbf{p}_\phi|/ m_\phi)$ of these boosted $\phi$ particles will decay on the typical $d\sim\unit[8]{kpc}$ 
journey from the center of our galaxy to us, leading to the antideuteron flux
\begin{align}
J_{\anti{d}} &\sim\frac{3 \times 10^{-5}}{\unit{m^2\, s\, sr}} \,\frac{t_\text{Universe}}{\tau_\chi}\,\left(\frac{\unit[10^7]{TeV}}{\Lambda}\right)^4 ,
\end{align}
with $E_{\anti{d}}\simeq\unit[40]{GeV}$. 
In addition to the galactic contribution there is an extragalactic $\phi$ flux of similar magnitude as $J_\phi$ in Eq.~\eqref{eq:J_phi_galaxy}, albeit with a red-shifted continuous spectrum~\cite{Esmaili:2012us}, shown in Fig.~\ref{fig:phiflux}. Being of extragalactic origin these $\phi$ have more time to decay and thus give the dominant antinucleon flux if $\phi$ is very long lived.

The resulting boosted $\anti{d}$ flux is large enough to be detectable in AMS as long as $\Lambda \lesssim \unit[10^7]{TeV}$ in this example, or larger if we increase $m_\phi$ or take into account the extragalactic flux.
In this scenario there are no competing astrophysical signatures such as a large antiproton flux; the only limiting factor of the antideuteron flux comes from the size of $\Lambda$, which can be constrained in a given UV-complete model, to be discussed below.

On a side note, the heavier DM particle $\chi$ could easily be envisioned to have 
annihilation channels into $b$-quarks or $\tau$ particles that can produce the  
$\gamma$-ray excesses observed in the galactic 
center~\cite{Hooper:2010mq,Hooper:2011ti,Abazajian:2012pn,TheFermi-LAT:2015kwa,Karwin:2016tsw} 
and Andromeda~\cite{Karwin:2019jpy}.

\subsection{UV completion}
\label{sec:deuteronUV}

As discussed above, the interaction of Eq.~\eqref{eq:lagrangianD} with an effective scale up to $\Lambda\sim \unit[10^{19}]{TeV}$ will produce upwards of $10^7$ non-relativistic antideuteron nuclei per year in our galaxy if $\phi$ itself is DM.
Producing $\phi$ instead from the decay of a heavier DM candidate can easily give rise to a \emph{boosted} observable $\anti{d}$ flux without other antimatter contributions, which however requires $\Lambda \lesssim \unit[10^7]{TeV}$.
By construction these antideuteron events come without a large flux of accompanying antiprotons, markedly different from the usual antideuteron production by DM via coalescence.

The required multi-TeV values for $\Lambda$ appear at first sight perfectly innocuous and well outside of most terrestrial experiments. However, a full UV-complete model has to be based on \emph{quark} couplings rather than nucleons, which increases the dimension of the underlying operator. Naively, a three-quark operator will hadronize into one nucleon $N$ via  $qqq\to \Lambda_\mathrm{QCD}^3 N$, which implies that the $\Lambda$ in our Eq.~\eqref{eq:lagrangianD} is related to a quark-level effective-field-theory scale $\Lambda_\mathrm{UV}$ via
\begin{align}
\frac{1}{\Lambda^2}\lesssim \frac{\Lambda_\mathrm{QCD}^6 }{\Lambda_\mathrm{UV}^{8}}
\simeq \frac{1}{(\unit[10^{11}]{TeV})^2} \left(\frac{\unit[1]{TeV}}{\Lambda_\mathrm{UV}}\right)^{8} .
\end{align}
$\Lambda_\mathrm{UV}$  is thus parametrically suppressed compared to $\Lambda$ and below 
$\Lambda_\mathrm{UV}\sim \unit[100]{TeV}$ if we want $\phi$ DM to produce non-relativistic antideuteron and even lower around
$\Lambda_\mathrm{UV}\sim \unit[100]{GeV}$ if we want \emph{boosted} $\phi$, using $\Lambda_\mathrm{QCD}\sim \unit[200]{MeV}$.  
It is this $\Lambda_\mathrm{UV}$ that is ultimately related to the masses of the integrated-out mediators, which should be below $4\pi \Lambda_\mathrm{UV}$ on account of perturbative unitarity. Since most of the mediators will be colored and charged we expect them to be subject to LHC searches for diquarks and leptoquarks, which typically yield lower limits for the masses around TeV. It has to be stressed that we are making very conservative estimates here with a overly simplified approach to hadronization.
Still, at this level of scrutiny it then seems that we can indeed build UV-complete collider-safe models that produce a significant amount of antideuteron in our galaxy.
It is more difficult to obtain large-enough fluxes of \emph{boosted} $\anti{d}$ from DM decay since the required new-physics scales lie in the range that is probed by the LHC. A proper comparison of collider constraints and antinuclei fluxes requires a full UV complete model and  as well as an improvement of our nuclear-physics calculations, which we leave for future work.

Having discussed the production of the simplest antinucleus above, we will now turn to discuss how heavier antinuclei can be produced in an analogous fashion, starting with antihelium $\aHe{3}$ in Sec.~\ref{sec:helium3} and $\aHe{4}$ in App.~\ref{sec:helium4}.

\section{Models for \texorpdfstring{$\aHe{3}$}{anti-He3} production}
\label{sec:helium3}

As an explicit realization of $\aHe{3}$ production
let us introduce a new SM-singlet complex scalar $\phi$ with baryon number $-3$ and 
lepton number $-1$ (the quantum numbers of the antitritium atom). 
We assume these 
symmetries to be sufficiently conserved so that they govern the $\phi$ decay modes, with one of the lowest-dimensional operators given by
\begin{align}
\mathcal{L} = \frac{\phi\, \overline{n}^c \gamma_5 {n}\, \overline{p}^c 
\gamma_5 {e}}{\Lambda^3} 	+ \text{h.c.}
\label{eq:lagrangianHe3}
\end{align}
In the non-relativistic limit this corresponds to a coupling 
$\phi\ {n}_\uparrow {n}_\downarrow {p}_\uparrow {e}_\downarrow$, 
$f_{\uparrow\downarrow}$ being the spin up/down states of particle $f$. Operators 
similar to Eq.~\eqref{eq:lagrangianHe3} but without the $\gamma_5$ would create neutrons 
in the same spin state and thus be suppressed by Fermi statistics in the 
non-relativistic limit; 
Eq.~\eqref{eq:lagrangianHe3} is then expected to be the dominant operator of interest for $\phi$ decay.
We stress again that baryon and lepton number are conserved in all interactions underlying Eq.~\eqref{eq:lagrangianHe3},
so the mediator particles will not induce proton decay or neutron-antineutron oscillations.

\subsection{Decay channels}

We will consider the situation where $m_\phi$ is close to the total
mass of the decay products, i.e.~$m_\phi\sim 2m_\mathrm{n} + m_\mathrm{p}$, neglecting again the electron mass.
In this situation, the nucleons can be treated as point particles and we can ignore
the underlying quark structure. 
The possible final states then depend strongly on $m_\phi$;
if $m_\phi$ is \emph{below} the threshold $2m_\mathrm{n}+m_\mathrm{p}$, the 
decay \emph{has} to go into a bound state of the nucleons in order to be energetically
 allowed. The relevant thresholds are given in Tab.~\ref{tab:thresholdsB3}, which shows that 
in the range $\unit[0.5]{MeV}<m_\phi-m_{\He{3}}< \unit[6.8]{MeV}$, the dominant kinematically
allowed decay induced by Eq.~\eqref{eq:lagrangianHe3} is $\phi \to \anti{t}\,\anti{e}$, followed by the beta decay of the antitritium nucleus $\anti{t}$ into $\aHe{3}$ within $\unit[12]{yr}$.\footnote{The \emph{direct} decay of $\phi$ into $\aHe{3}$ is further suppressed by  $G_\mathrm{F}$ and will not be discussed here.}
 This is then a source for production of $\aHe{3}$ as long as the lifetime for
$\phi$ is not too long, which can indeed be the case despite the small available phase 
space, as we will now show.

\begin{table}[t]
	\begin{tabular}{lc}
		state & $m-m_{\He{3}}$ [MeV] \\
		\hline
$\mathrm{t}^+$	& 0.5\\
$\mathrm{d}^+\, \mathrm{n}$		& 6.8 \\
$\mathrm{p}\, \mathrm{n}\,\mathrm{n}$		& 9.0 \\
\hline

	\end{tabular}
	\caption{Lowest lying states with baryon number~3 and electric charge 1, with mass 
	$m$ relative to that of $\He{3}^{+}$, $m_{\He{3}}= \unit[2808.4]{MeV}$. The  deuteron and triton nuclei are as usual denoted as 
	d$^+$ and t$^+$ instead of ${}^2\mathrm{H}^+$ and ${}^3\mathrm{H}^+$. We will drop the
	ionization superscripts in the following since we always refer to nuclei. 
	}
	\label{tab:thresholdsB3}
\end{table}

Assuming $m_\phi$  to be close to $m_{ \He{3}}$ ensures that the nucleons and nuclei 
involved in the process are non-relativistic. 
It is then convenient to calculate the decay rate non-relativistically with 
Fermi's golden rule,
\begin{align}
\Gamma(\phi \to \anti{t} \, \anti{e}) = 2\pi |V(\phi \to \anti{t} \, \anti{e})|^2 \rho \,,
\end{align}
$\rho = (m_\phi-m_\mathrm{t})^2/(2\pi^2)$ being the phase-space density. 
$V(\phi \to \anti{t} \, \anti{e})$ is
the appropriate transition element, here approximated as
 \begin{align}
 \begin{split}
\left|V(\phi \to \anti{t} \, \anti{e})\right|^2 &\simeq \frac{\sum_\text{spins} 
|\mathcal{M}(\phi \to \anti{n}\,\anti{n}\,\anti{p}\,\anti{e}) |^2}
{2m_\phi \, 2 E_{\mathrm{n}_1}\, 2 E_{\mathrm{n}_2}\, 2 E_{\mathrm{p}}\, 2 E_{\mathrm{e}}}\, |\langle \anti{n}\,\anti{n}\,\anti{p} | \anti{t} \rangle|^2 \,,
\end{split}
\end{align}
where the first factor is the 
hard process and the second factor is  the overlap of the nucleon wave 
function with the tritium nucleus.
The  hard process $\phi (k) \to \anti{n}(k_{\mathrm{n}_1}) \anti{n}(k_{\mathrm{n}_2}) \anti{p}
(k_{\mathrm{p}}) \anti{e}
(k_{\mathrm{e}})$
has a  squared matrix element 
\begin{align}
\sum_\text{spins} |\mathcal{M}(\phi \to \anti{n}\,\anti{n}\,\anti{p}\,
\anti{e})|^2 \simeq \frac{ 32}
{\Lambda^{6}}(k_{\mathrm{n}_1}k_{\mathrm{n}_2} + m_\mathrm{n}^2)(k_{\mathrm{p}} k_{\mathrm{e}}) \,.
\label{eq:msq3}
\end{align}
The overlap $\langle \anti{n}\,\anti{n}\,\anti{p} | \anti{t} \rangle$ of the 
nucleons with the tritium nucleus wave function can be estimated
by treating the nucleons as moving in a mean field potential. Once the center of
mass motion is factored out, the nucleus is described by a product of
two wave functions $\psi(x_i-x_j)$ for the two relative coordinates. Since the
decay of $\phi$ is assumed to produce the three nucleons initially at a single point, we 
can estimate the  overlap as
 $|\langle \anti{n}\,\anti{n}\,\anti{p} | \anti{t} \rangle|^2\simeq (|\psi (0)|^2)^2$.
In simple shell models, the wave function $\psi (0)$ can be approximately related to 
the radius of t, $|\psi (0)|\sim r_{\mathrm{t}}^{-3/2}$, or simply 
$|\psi (0)|\sim (\unit[100]{MeV})^{3/2}$ as a nuclear-physics energy scale.
A more accurate calculation of this matrix element is desirable and will be left to
future work.

\begin{figure}[t]
\includegraphics[width=0.48\textwidth]{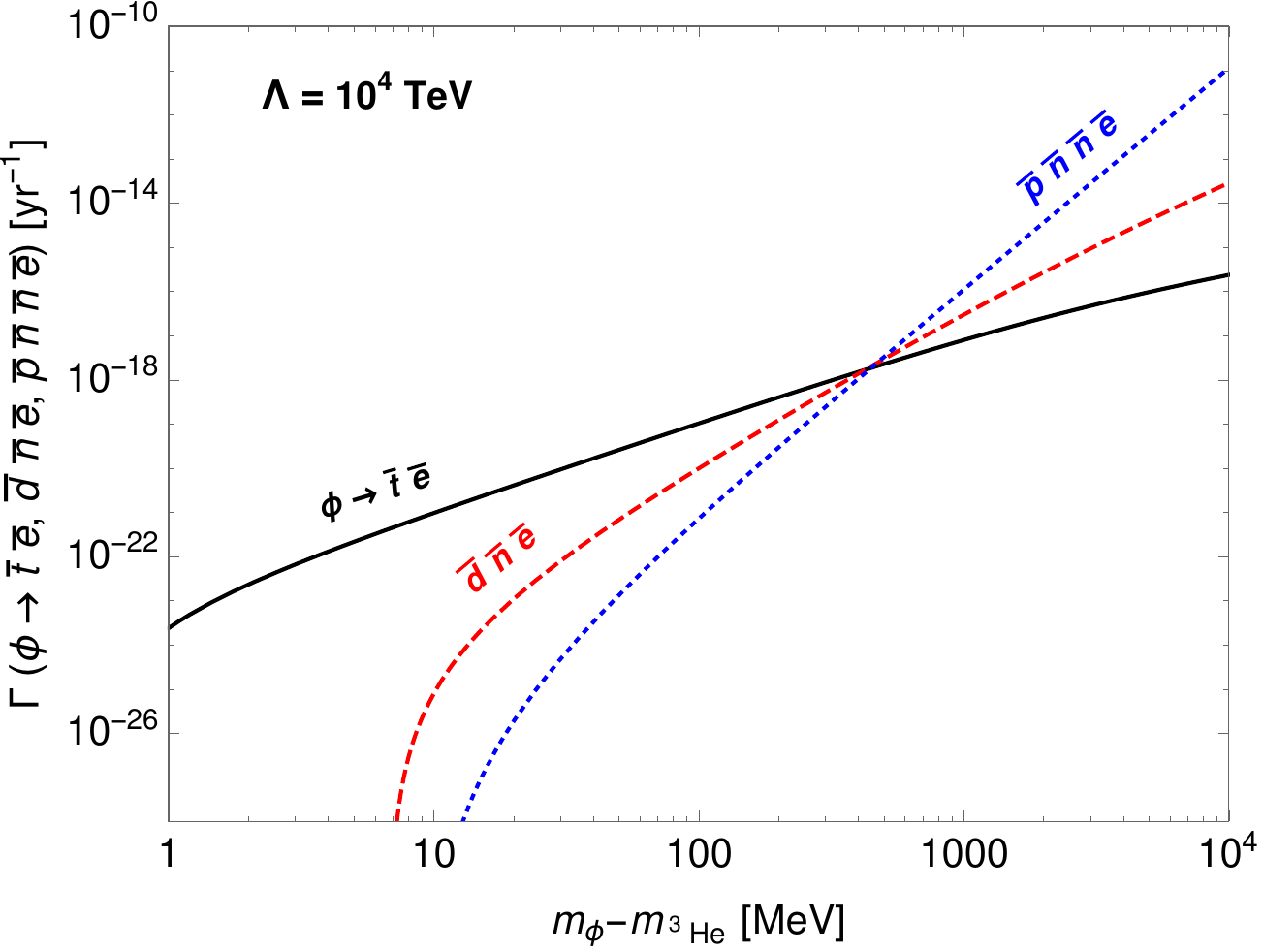}
\caption{Decay rates of $\phi$ to $\anti{t} \,\anti{e}$, $\anti{d} \,\anti{n} \,\anti{e}$, and  $\anti{p} \,\anti{n} \,\anti{n} \,\anti{e}$ for $\Lambda=\unit[10^4]{TeV}$.
We again denote $\mathrm{d}\equiv {}^2\mathrm{H}^+$ and $\mathrm{t} \equiv {}^3\mathrm{H}^+$.
All nuclear matrix elements are taken to be $\unit[100]{MeV}$ to the appropriate power.
The rates are calculated assuming $m_\phi-m_{\He{3}}\ll m_{\He{3}}$.
}
\label{fig:lifetimeHe3}
\end{figure}

The nucleons are non-relativistic but the electrons relativistic, leading to the decay rate 
\begin{align}
\Gamma(\phi \to \anti{t} \, \anti{e}) &\simeq \frac{ 2 (m_\phi-m_\mathrm{t})^2}
{\pi\,m_\phi \,\Lambda^{6}} |\psi (0)|^4 \,,
\label{eq:decay_to_t}
\end{align}  
which we expect to be valid for $m_\phi-m_\mathrm{t}< m_\mathrm{t}$.
With $|\psi (0)|^4\simeq (\unit[100]{MeV})^{6}$ this yields the lifetime 
\begin{align}
\tau &\simeq \unit[10^{6}]{Gyr} \left(\frac{\unit[10]{MeV}}{m_\phi-m_{\He{3}}}\right)^2
\left(\frac{\Lambda}{\unit[10^3]{TeV}}\right)^{6} ,
\end{align}
illustrated in Fig.~\ref{fig:lifetimeHe3},
which can easily surpass the age of the universe, 
$t_\text{Universe}\simeq \unit[14]{Gyr}$.

For masses $m_\phi-m_{\He{3}} > \unit[7]{MeV}$, the decay channel $\phi \to \anti{d}\,\anti{n}\,\anti{e}$ into antideuterons opens up, which we calculate in analogy to before using Fermi's golden rule. The nuclear matrix element $|\langle \anti{n}\,\anti{p} | \anti{d} \rangle|^2\simeq |\psi (0)|^2$ is assumed to be $(\unit[100]{MeV})^3$ in our numerical evaluations. For slow hadrons the decay rate then reads
\begin{align}
\begin{split}
\Gamma(\phi \to \anti{d} \, \anti{n} \, \anti{e}) &\simeq \frac{16}{105\pi^3 m_\phi \Lambda^6}  
\frac{m_\mathrm{d}^{3/2} m_\mathrm{n}^{3/2}}{(m_\mathrm{d}+m_\mathrm{n})^{3/2}} \\
&\quad \times \left(m_\phi - m_\mathrm{d}-m_\mathrm{n}\right)^{7/2} |\psi (0)|^2 \,,
\end{split}
\label{eq:decay_to_dne}
\end{align}  
which starts to dominate over the two-body decay for $m_\phi-m_{\He{3}} \gtrsim \unit[500]{MeV}$ (Fig.~\ref{fig:lifetimeHe3}).

Finally, when $m_\phi$ is above the threshold for decay into free nucleons, 
$m_\phi >2m_\mathrm{n} + m_\mathrm{p}$, but still small enough to not resolve 
the underlying quark structure or heavier baryons such as $\Sigma$ and $\Delta$, 
$\phi$ will decay into free nucleons.
Using the matrix element from Eq.~\eqref{eq:msq3} we find the decay rate  near threshold
\begin{align}
\begin{split}
\Gamma(\phi \to  \anti{n}\,\anti{n}\,\anti{p}\,\anti{e}) 
&\simeq \frac{137}{10468 \pi^5 \Lambda^{6}} \frac{m_\mathrm{n}^3 m_\mathrm{p}^{3/2}}{\left(2m_\mathrm{n}+m_\mathrm{p} \right)^{5/2}} \\
&\quad \times\left(m_\phi -2m_\mathrm{n}-m_\mathrm{p} \right)^{5} .
\end{split}	
\end{align}
This decay to free nucleons dominates for large $m_\phi$ and  will eventually 
lead to constraints from antiproton-flux measurements that plague most DM models that aim to produce heavier antinuclei~\cite{Carlson:2014ssa,Cirelli:2014qia,Coogan:2017pwt}.

The main conclusion from this analysis is that $\phi$ decays can produce 
almost exclusively $\aHe{3}$ for $0\lesssim m_\phi-m_{\He{3}} \ll \unit[100]{MeV}$ (see Fig.~\ref{fig:lifetimeHe3}), while for 
$m_\phi-m_{\He{3}} \sim \unit[500]{MeV}$ we expect 
similar amounts of $\aHe{3}$, $\anti{d}$, and $\anti{p}$. For even 
larger $\phi$ masses the $\aHe{3}$ fraction will decrease and the 
amount of $\anti{p}$ will become significant, eventually leading back 
to coalescence fractions for the antinuclei.
In order to dominantly produce $\anti{He}$ we therefore need $m_\phi \simeq \unit[3]{GeV}$.\footnote{An actual comparison of the various antimatter channels has to include the spectral information, so it is entirely possible to also consider the region $m_\phi \gg \unit[3]{GeV}$, which will not be done here.}

\subsection{\texorpdfstring{$\phi$}{phi} as dark matter}
\label{sec:phiDM}

Since the total lifetime of the new particle $\phi$ can be  longer than the 
age of the universe, it is conceivable that $\phi$ is DM or a subcomponent of it. 
The appropriate relic density
could be produced 
through interactions like the Higgs portal 
$|\phi|^2 |H|^2$ or new gauge interactions,
which allow for $\phi$ pair production without affecting its 
decay. Alternatively, as a result of the assigned baryon number it is reasonable 
to expect $\phi$ to have an \emph{asymmetry} similar to the baryon asymmetry, which 
is consistent with the $\sim\unit[3]{GeV}$ DM mass~\cite{Zurek:2013wia}.
 
Assuming $\phi$ to be DM, we can in principle calculate the antinuclei flux.
As a conservative lower limit on the lifetime we take $\tau > \unit[10^{25}]{s}$~\cite{Slatyer:2016qyl} in order to suppress energy injection during CMB formation from the fast positron that is emitted in $\phi$ decays. For $m_\phi =\unit[3]{GeV}$ this corresponds to a lower limit $\Lambda >\unit[7\times 10^3]{TeV}$.
 There are about 
$10^{12}M_\odot/(\unit[3]{GeV})\simeq 4\times 10^{68}$ DM particles in our galaxy, a fraction
$\unit{yr}/\tau < 3 \times 10^{-18}$ of which decay into $\aHe{3}$ per year.
Even with such a long lifetime we can thus produce up to $10^{51}$ $\aHe{3}$ nuclei per year in this scenario.
Pushing $\Lambda \to \unit[ 10^{11}]{TeV}$ still yields $\mathcal{O}(10^8)$ antihelium nuclei per year within the Milky Way, with negligible numbers of antiproton and antideuteron.
Producing a significant number of antihelium is thus not difficult in our model.

However, all of these $\aHe{3}$ are by construction non-relativistic, with 
 kinetic energy $(m_\phi-m_{\He{3}})^2/(2 m_{\He{3}})= \unit[6.5]{MeV}$, far below the 
kinetic-energy threshold for AMS detection of around GeV/nucleon. 
Still, a small fraction of these antinuclei might be accelerated 
by astrophysical processes such as supernovae shock waves, leading to a relativistic 
flux of antihelium that could explain the preliminary AMS events. Most of the remaining 
non-relativistic antinuclei will eventually come in contact with normal 
matter and annihilate, giving rise to characteristic photon spectra that could be detected 
by the Fermi-LAT telescope. A  determination of this fraction is necessary to evaluate whether our model is better constrained by AMS or Fermi-LAT.
Note that we ignored induced-decay processes such as $\mathrm{e}\, \phi\to  \anti{t}\,\gamma$ that could increase the total number of $\aHe{}$ and even boost them, but require a dedicated astrophysical simulation.

$\phi$ could also be detectable in 
large underground detectors such as~Super-Kamiokande
through exotic events of the form 
$\phi\,\mathrm{p}\to  \anti{n}\,\anti{n}\,\anti{e}$.
This process can be interpreted as an effective proton lifetime, albeit with an energy release $E > m_\mathrm{p}$, that is determined by the interaction rate~\cite{Davoudiasl:2010am,Davoudiasl:2011fj}
\begin{align}
\tau^\text{eff}_\mathrm{p} = \frac{1}{n_\phi\, \sigma v (\phi\, \mathrm{p}\to  \anti{n}\,\anti{n}\,\anti{e})} \,,
\end{align}
with DM number density at Earth $n_\phi = \rho_\phi/m_\phi \sim 0.1/\unit{cm^3}$, and $\sigma v $ the cross section between the non-relativistic $\phi$ and proton. We calculate this cross section with \texttt{CalcHEP}~\cite{Belyaev:2012qa} for $m_\phi =\unit[3]{GeV}$, ignoring formation of final state bound states for simplicity, as
\begin{align}
\hspace{-1.1ex}\sigma v (\phi\, \mathrm{p}\to  \anti{n}\,\anti{n}\,\anti{e})
\simeq 3\times 10^{-61}\,\frac{\unit{cm^3}}{\unit{s}}\left(\frac{\unit[7\times 10^3]{TeV}}{\Lambda}\right)^{6},
\end{align}
which yields an effective proton lifetime $\tau^\text{eff}_\mathrm{p} $ above $\unit[10^{54}]{yr}$. This is comfortably  larger than typical proton lifetime limits, but of course no dedicated search for this dramatic process exists.

\subsection{\texorpdfstring{$\phi$}{phi} produced by dark matter}
\label{sec:helium3subDM}

In order to not rely on an astrophysical $\aHe{3}$ acceleration mechanism we can imagine $\phi$ to be \emph{produced boosted} by a heavier DM particle $\chi$, either via annihilations or decays (e.g.~$\chi \to \phi\phi$ if 
$\chi$ carries baryon number $-6$ or $\chi \to \phi\phi\phi$ if $\chi$ has baryon number $-9$).
In this basic setup there are no CMB constraints on the lifetimes of $\chi$ or $\phi$, other than $\tau_\chi>t_\text{Universe}$ by assumption.
In particular, the decay rate of $\phi$ is only constrained by the AMS antihelium flux and UV-considerations for $\Lambda$, which we address later.

Consider for example $\chi\to\phi\phi$ in our galaxy with $m_\chi =\unit[80]{GeV}$ and $m_\phi =\unit[3]{GeV}$, which gives a flux of $\phi$ particles with $E_\phi=\unit[40]{GeV}$  at Earth of
\begin{align}
J_\phi &= \frac{2}{4\pi m_\chi \tau_\chi}\int_0^\infty \text{d} s \,\rho_\text{Halo}[r(s)]\sim\frac{10^6}{\unit{m^2\, s\, sr}} \,\frac{t_\text{Universe}}{\tau_\chi} \,,
\label{eq:J_phi}
\end{align}
where the integral is over the line of sight~\cite{Esmaili:2012us}. 
Only a small fraction $d/(\tau\, |\mathbf{p}_\phi|/ m_\phi)$ of these boosted $\phi$ particles will decay on the typical $d\sim\unit[8]{kpc}$ 
journey from the center of our galaxy to us, leading to the antihelium flux
\begin{align}
J_{\aHe{3}} &\sim\frac{7\times 10^{-4}}{\unit{m^2\, s\, sr}} \,\frac{t_\text{Universe}}{\tau_\chi}\,\left(\frac{\unit[10^3]{TeV}}{\Lambda}\right)^6 ,
\end{align}
with $E_{\aHe{3}}\simeq\unit[40]{GeV}$. 
In addition to the galactic contribution there is an extragalactic $\phi$ flux of similar magnitude as $J_\phi$ in Eq.~\eqref{eq:J_phi}, albeit with a red-shifted continuous spectrum~\cite{Esmaili:2012us}, shown in Fig.~\ref{fig:phiflux}. Being of extragalactic origin these $\phi$ have more time to decay and thus give the dominant antinucleon flux if $\phi$ is very long lived.

The resulting boosted $\anti{He}$ flux is large enough to be detectable in AMS as long as $\Lambda \lesssim \unit[10^4]{TeV}$ in this example, or larger if we increase $m_\phi$. 
In this scenario there are no competing astrophysical signatures such as a large antiproton flux; the only limiting factor of the antihelium flux comes from the size of $\Lambda$, which can be constrained in a given UV-complete model, to be discussed below.

On a side note, the heavier DM particle $\chi$ could easily be envisioned to have 
annihilation channels into $b$-quarks or $\tau$ particles that can produce the  
$\gamma$-ray excesses observed in the galactic 
center~\cite{Hooper:2010mq,Hooper:2011ti,Abazajian:2012pn,TheFermi-LAT:2015kwa,Karwin:2016tsw} 
and Andromeda~\cite{Karwin:2019jpy}.

\subsection{UV completion}
\label{sec:UV}

As discussed above, the interaction of Eq.~\eqref{eq:lagrangianHe3} with an effective scale up to $\Lambda\sim \unit[10^{11}]{TeV}$ will produce upwards of $10^8$ non-relativistic antihelium nuclei in our galaxy if $\phi$ itself is DM.
Producing $\phi$ instead from the decay of a heavier DM candidate can easily give rise to a \emph{boosted} observable $\aHe{3}$ flux without other antimatter contributions, which however requires $\Lambda \lesssim \unit[10^4]{TeV}$.
By construction these antihelium events come without a large flux of accompanying antiprotons or antideuterons, markedly different from the usual antihelium production by DM via coalescence.

The required multi-TeV values for $\Lambda$ appear at first sight perfectly innocuous and well outside of most terrestrial experiments. However, a full UV-complete model has to be based on \emph{quark} couplings rather than nucleons, which increases the dimension of the underlying operator. Naively, a three-quark operator will hadronize into one nucleon $N$ via  $qqq\to \Lambda_\mathrm{QCD}^3 N$, which implies that the $\Lambda$ in our Eq.~\eqref{eq:lagrangianHe3} is related to a quark-level effective-field-theory scale $\Lambda_\mathrm{UV}$ via
\begin{align}
\frac{1}{\Lambda^3}\lesssim \frac{\Lambda_\mathrm{QCD}^9 }{\Lambda_\mathrm{UV}^{12}}
\simeq \frac{1}{(\unit[10^{11}]{TeV})^3} \left(\frac{\unit[1]{TeV}}{\Lambda_\mathrm{UV}}\right)^{12} .
\end{align}
$\Lambda_\mathrm{UV}$  is thus parametrically suppressed compared to $\Lambda$ and below 
$\Lambda_\mathrm{UV}\sim \unit[1]{TeV}$ if we want $\phi$ DM to produce non-relativistic antihelium and even lower around
$\Lambda_\mathrm{UV}\sim 10$--$\unit[30]{GeV}$ if we want \emph{boosted} $\phi$, using $\Lambda_\mathrm{QCD}\sim \unit[200]{MeV}$.  
It is this $\Lambda_\mathrm{UV}$ that is ultimately related to the masses of the integrated-out mediators, which should be below $4\pi \Lambda_\mathrm{UV}$ on account of perturbative unitarity. Since most of the mediators will be colored and charged we expect them to be subject to LHC searches for diquarks and leptoquarks, which typically yield lower limits for the masses around TeV. It has to be stressed that we are making very conservative estimates here with a overly simplified approach to hadronization.
Still, at this level of scrutiny it then seems that we can indeed build UV-complete collider-safe models that produce a significant amount of antihelium in our galaxy, but it seems difficult to obtain large enough fluxes of \emph{boosted} $\phi$ from DM decay.

\begin{figure}[t]
\includegraphics[width=0.48\textwidth]{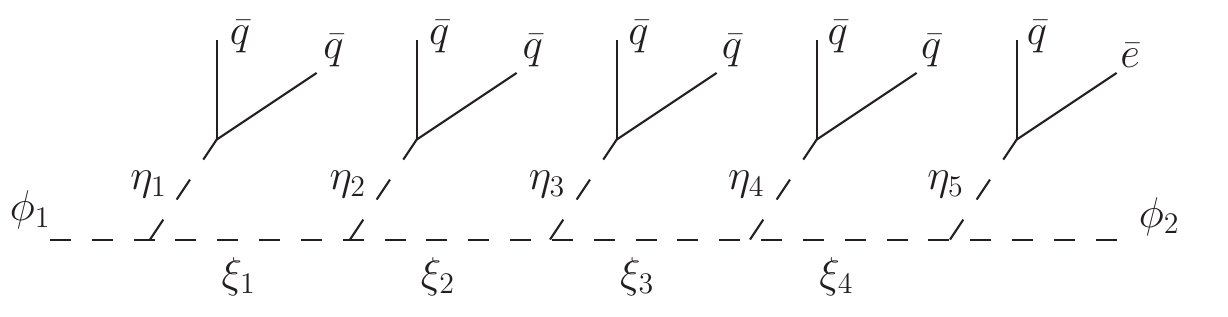}
\caption{UV-realization of a relevant operator for $\phi_1\to\phi_2\, \anti{t}\,\anti{e}$.
}
\label{fig:resonant_decay_chain}
\end{figure}

Motivated by this result let us consider variations of our model that do not suffer from this potentially dangerous UV suppression. Instead of $\phi \to \anti{t}\,\anti{e}$ consider $\phi_1\to\phi_2\, \anti{t}\,\anti{e}$, again close to phase-space closure $m_{\phi_1} \sim  m_t +m_{\phi_2}$ in order to suppress antiproton production. The discussion is essentially analogous to before upon replacing $m_\phi \to m_{\phi_1}-m_{\phi_2}$ but now we are able to push $m_{\phi_j}$ to the TeV scale. Assuming all new particles, including the colored mediators, to be around the TeV scale makes it possible to enhance the overall $\phi_1\to\phi_2\, \anti{t}\,\anti{e}$ via resonances, i.e.~nearly on-shell mediator particles. This is illustrated in Fig.~\ref{fig:resonant_decay_chain}, which shows a UV-complete origin of the nine-quark operator relevant for $\phi_1\to\phi_2\, \anti{t}\,\anti{e}$. Here, the scalars $\xi_j$ can lead to a resonant enhancement if their masses are close to $m_{\phi_1}$ because the fermions only carry away small amounts of energy compared to $m_{\phi_1}$. This can compensate for the unwelcome $\Lambda_\mathrm{QCD}^9$ suppression that is generic for nine-quark operators and makes it possible to have UV-complete realizations of heavy antimatter production that are consistent with collider constraints.
A calculation of $\phi_1\to\phi_2\, \anti{t}\,\anti{e}$ that takes into account the quark hadronization and subsequent nucleus formation involving such resonant effects is clearly non-trivial and will not be attempted here.

\section{Conclusions}
\label{sec:conclusions}

Standard astrophysical models and even most DM models predict antimatter fluxes that 
are strongly suppressed for antinuclei with $A\geq 2$, making it very difficult (impossible) to obtain $\aHe{3}$ ($\aHe{4}$) fluxes that could be detected in current and upcoming experiments such as AMS-02.
The strong hierarchy of antiproton to heavier antinuclei fluxes rests on coalescence models which quantify the probability of forming antinuclei in a given high-energy process.

We have shown here that it is possible to circumvent the predicted coalescence ratios of antinuclei fluxes by considering \emph{low-energy} production processes, for which the nuclear binding energy and phase space considerations play important roles. It is, for example, possible to
introduce new particles $\phi$ that dominantly decay into $\anti{d}$ or $\anti{He}$ without the typical accompanying larger antiproton flux. The key observation is that the desired number of final state nucleons can be enforced by assigning a baryon number to $\phi$ while nuclei formation can be enforced by reducing the available phase space.
In the simplest realization of this idea $\phi$ itself forms DM and slowly decays into non-relativistic antinuclei, resembling to some degree antimatter regions. Most of these antinuclei will come into contact with normal matter and produce characteristic photon signals, while a small number of antihelium nuclei might be accelerated towards Earth by astrophysical processes.
Without relying on such an acceleration mechanism we have also discussed a variation of this model in which $\phi$ itself is produced boosted by a heavier DM candidate, increasing the flux of relativistic antinuclei. However, a sufficiently small lifetime of $\phi$ in this case seems to be difficult to achieve in UV-complete models for antinuclei heavier than $\anti{d}$. Interestingly, the limiting factor of antideuteron or antihelium production in this model is not the amount of lighter antinuclei as in other DM models but rather collider constraints on the mediator particles.

Future work could proceed in various directions.
The UV-complete models have to be studied to evaluate the signatures of the underlying colored mediator particles at the LHC.
Improved nuclear-physics calculations are necessary to pin down the relevant matrix elements that determine the fluxes and branching ratios of $\aHe{4}$, $\aHe{3}$, $\anti{d}$, and $\anti{p}$. Dedicated astrophysical simulations for antinuclei propagation are required to derive the final antimatter fluxes and compare them to indirect constraints e.g.~from gamma-ray searches.

\section*{Acknowledgements}
We thank Tim Tait, Manoj Kaplinghat, and Mauro Valli for discussions and comments on the manuscript.
This work was supported by the National Science Foundation under Grant No.~PHY-1620638.
J.H.~is further supported by a Feodor Lynen Research Fellowship of the Alexander von Humboldt Foundation.

\appendix

\section{Models for \texorpdfstring{$\aHe{4}$}{anti-He4} production}
\label{sec:helium4}

In the main text we have discussed simple ways to produce $\anti{d}$ or $\aHe{3}$ without accompanying $\anti{p}$. The same idea can be applied to other antimatter, $\aHe{4}$ being particularly intriguing considering the potential observation of two events in AMS~\cite{ams_talk_2019}. 
As an explicit realization of $\aHe{4}$ production in complete analogy to Sec.~\ref{sec:helium3}
let us introduce a new SM-singlet complex scalar $\phi$ with baryon number $-4$ and 
lepton number $-2$ (the quantum numbers of the $\aHe{4}$ atom). 
We assume these 
symmetries to be sufficiently conserved so that they govern the $\phi$ decay modes, with one of the lowest-dimensional operators given by
\begin{align}
\mathcal{L} = \frac{1}{(2\Lambda)^6} \phi\, \overline{n}^c \gamma_5 {n}\, \overline{p}^c 
\gamma_5 {p}	 \, \overline{e}^c \gamma_5 {e} + \text{h.c.}
\label{eq:lagrangian}
\end{align}
In the non-relativistic limit this corresponds to a coupling 
$\phi\ {n}_\uparrow {n}_\downarrow {p}_\uparrow {p}_\downarrow {e}_\uparrow {e}_\downarrow$, 
$f_{\uparrow\downarrow}$ being the spin up/down states of particle $f$. Operators 
similar to Eq.~\eqref{eq:lagrangian} but without the $\gamma_5$ would create fermions 
in the same spin state and thus be suppressed by Fermi statistics in the 
non-relativistic limit; operators with more than 
two fermions of the same type, e.g.~$\overline{n}^c (\gamma_5) {n}\, 
\overline{n}^c (\gamma_5) {n}	 \, \overline{\nu}^c (\gamma_5) {\nu}$ will be suppressed as well.
Eq.~\eqref{eq:lagrangian} is then expected to be the dominant operator of interest for $\phi$ decay.

\subsection{Decay channels}

As in the previous case, we will consider the situation where $m_\phi$ is close to the total
mass of the decay products, i.e.~$m_\phi\sim 2m_\mathrm{n} + 2m_\mathrm{p}$, neglecting again the electron mass $m_\mathrm{e}$ for simplicity.
The nucleons can be treated as point particles and we can ignore
the underlying quark structure. 
Again, if $m_\phi$ is \emph{below} the threshold $2m_\mathrm{n}+2m_\mathrm{p}$, the 
decay \emph{has} to go into a bound state of the nucleons in order to be energetically
 allowed. The relevant thresholds are given in Tab.~\ref{tab:thresholds}, which shows that 
in the range $0<m_\phi-m_{\He{4}}< \unit[19.81]{MeV}$, the only kinematically
allowed decay induced by Eq.~\eqref{eq:lagrangian} is $\phi \to \aHe{4}\,\anti{e}\,\anti{e}$.
 This is then a source for production of $\aHe{4}$ as long as the lifetime for
$\phi$ is not too long. This can indeed be the case despite the small available phase 
space, as we will now show.

\begin{table}[t]
	\begin{tabular}{lc}
		state & $m-m_{\He{4}}$ [MeV] \\
		\hline
$\He{4}^{++}$	& 0\\
$\text{t}^{+}\,\mathrm{p}$ 		& 19.81\\
$\He{3}^{++}\,\mathrm{n}$ 	& 20.58\\
$\mathrm{d}^+\,\mathrm{d}^+$ 		& 23.80\\
$\mathrm{d}^+\, \mathrm{n}\, \mathrm{p}$		& 26.07\\
$\mathrm{n}\,\mathrm{n}\,\mathrm{p}\,\mathrm{p}$		& 28.30\\
\hline

	\end{tabular}
	\caption{Lowest lying ground states with baryon number 4 and electric charge 2, with mass 
	$m$ relative to $\He{4}^{++}$. The  deuteron and triton nuclei are as usual denoted as 
	d$^+$ and t$^+$ instead of ${}^2\mathrm{H}^+$ and ${}^3\mathrm{H}^+$. We will drop the
	ionization superscripts in the following. Not shown are excited states, which are 
	instead given in Tab.~\ref{tab:He4_states}.
	}
	\label{tab:thresholds}
\end{table}

\begin{table}[t]
	\begin{tabular}{cccc}
		$E-m_{\He{4}}$ [MeV] \,\, & $J^\pi$ \, & $\Gamma$ [MeV] \, & main decay\\
		\hline
		0	 	&	$0^+$	&	-- 		& --\\
		20.21 	&	$0^+$	&	0.50 	& t\,p\\
		21.01 	&	$0^-$	&	0.84 	& t\,p\\
		21.84 	&	$2^-$	&	2.01 	& t\,p\\
		23.33 	&	$2^-$	&	5.01 	& t\,p,$\He{3}$\,n\\
		23.64 	&	$1^-$	&	6.20	& t\,p,$\He{3}$\,n\\
		24.25 	&	$1^-$	&	6.10	& t\,p,$\He{3}$\,n\\
		25.28 	&	$0^-$	&	7.97	& t\,p,$\He{3}$\,n\\
		25.95 	&	$1^-$	&	12.66	& t\,p,$\He{3}$\,n\\
		27.42 	&	$2^+$	&	8.69	& d\,d\\
		28.31 	&	$1^+$	&	9.89	& t\,p,$\He{3}$\,n\\
		28.37 	&	$1^-$	&	3.92	& d\,d\\
		28.39 	&	$2^-$	&	8.75	& d\,d\\
		28.64 	&	$0^-$	&	4.89	& d\,d\\
		28.67 	&	$2^+$	&	3.78	& d\,d\\
		29.89 	&	$2^+$	&	9.72	& d\,d\\
\hline

	\end{tabular}
	\caption{$\He{4}$ states with energy relative to the ground state, $J^\pi$ quantum numbers, 
	decay width, and dominant decay channel. Adopted from Ref.~\cite{Tilley:1992zz}.
	}
	\label{tab:He4_states}
\end{table}

We calculate the decay rate non-relativistically with 
Fermi's golden rule,
\begin{align}
\Gamma(\phi \to \aHe{4} \, \anti{e}\, \anti{e}) = 2\pi |V(\phi \to \aHe{4} \, \anti{e} \,
\anti{e})|^2 \rho \,,
\end{align}
$\rho = (m_\phi-m_{\He{4}})^5/(120\pi^4)$ being the phase-space density, well-known from beta decays. 
$V(\phi \to \aHe{4} \, \anti{e} \,\anti{e})$ is
the appropriate transition element, here approximated as
 \begin{align}
 \begin{split}
\left|V(\phi \to \aHe{4} \,\anti{e}\, \anti{e})\right|^2 &\simeq \frac{\sum_\text{spins} 
|\mathcal{M}(\phi \to \anti{n}\,\anti{n}\,\anti{p}\,\anti{p}\,\anti{e}\,\anti{e}) |^2}
{2m_\phi \prod_{i=1,2}\prod_{j=\mathrm{n},\mathrm{p},\mathrm{e}} 2 E_{j_i}} \\
&\quad\times |\langle \anti{n}\,\anti{n}\,\anti{p}\,\anti{p} | \aHe{4} \rangle|^2 \,,
\end{split}
\end{align}
where the first factor is the 
hard process and the second factor is  the overlap of the nucleon wave 
function with the helium nucleus.
The  hard process $\phi (k) \to \anti{n}(k_{\mathrm{n}_1}) \anti{n}(k_{\mathrm{n}_2}) \anti{p}
(k_{\mathrm{p}_1}) \anti{p}(k_{\mathrm{p}_2}) \anti{e}(k_{\mathrm{e}_1}) \anti{e}
(k_{\mathrm{e}_2})$
has a  squared matrix element 
\begin{align}
\sum_\text{spins} |\mathcal{M}(\phi \to \anti{n}\,\anti{n}\,\anti{p}\,\anti{p}\,\anti{e}\,
\anti{e})|^2 = \prod_{j=\mathrm{n},\mathrm{p},\mathrm{e}} \frac{ k_{j_1}k_{j_2} + m_j^2}
{\Lambda^{4}} \,.
\label{eq:msq}
\end{align}
The overlap $\langle \anti{n}\,\anti{n}\,\anti{p}\,\anti{p} | \aHe{4} \rangle$ of the 
nucleons with the helium nucleus wave function can be estimated
by treating the nucleons as moving in a mean field potential. Once the center of
mass motion is factored out, the nucleus is described by a product of
three wave functions $\psi(x_i-x_j)$ for the three relative coordinates. Since the
decay of $\phi$ is assumed to produce the four nucleons initially at a single point, we 
can estimate the  overlap as
 $|\langle \anti{n}\,\anti{n}\,\anti{p}\,\anti{p} | \aHe{4} \rangle|^2\simeq (|\psi (0)|^2)^3$.
As before, we take $|\psi (0)|\sim (\unit[100]{MeV})^{3/2}$ as a nuclear-physics energy scale.
A more accurate calculation of this matrix element is desirable and will be left to
future work.

\begin{figure}[t]
\includegraphics[width=0.48\textwidth]{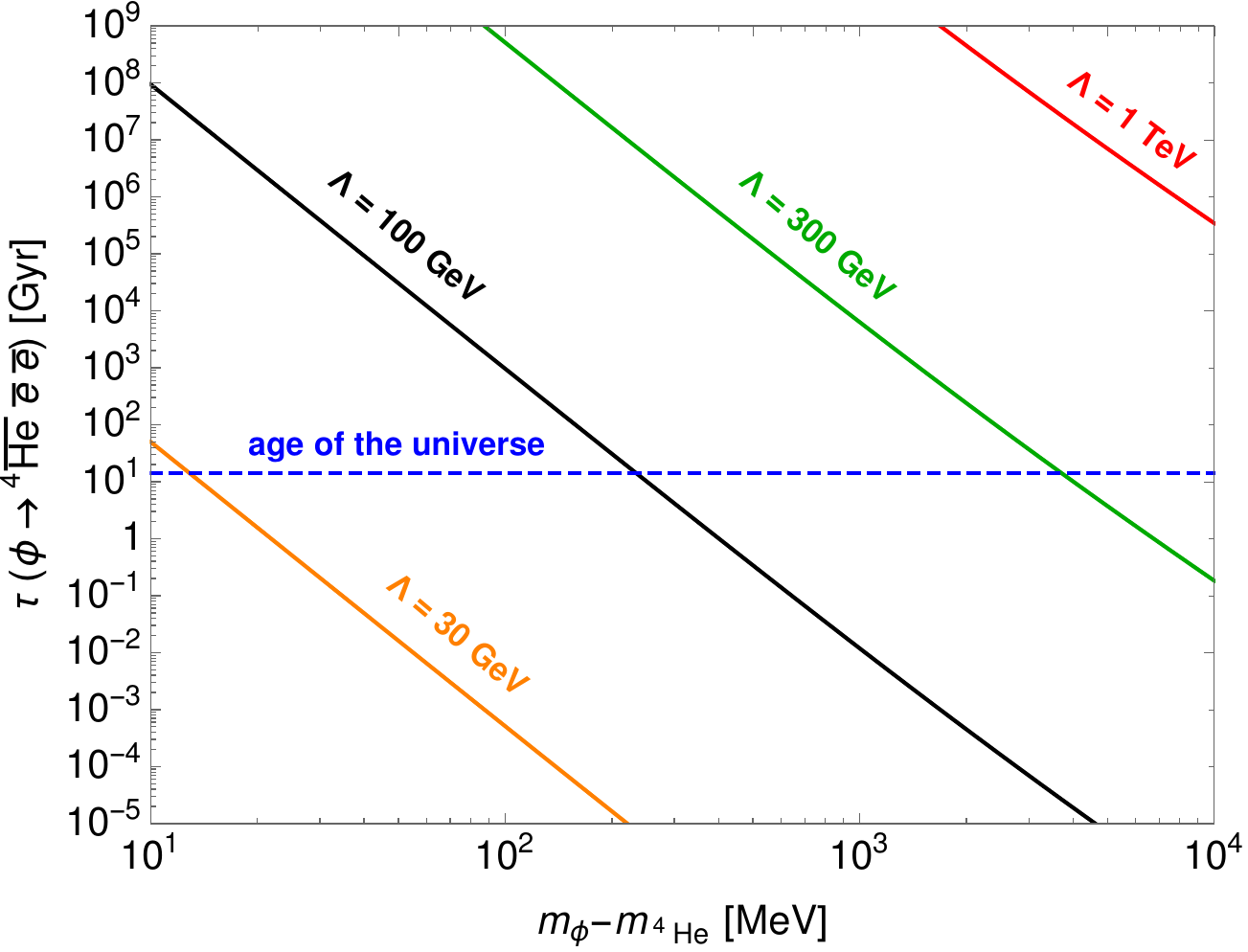}
\includegraphics[width=0.48\textwidth]{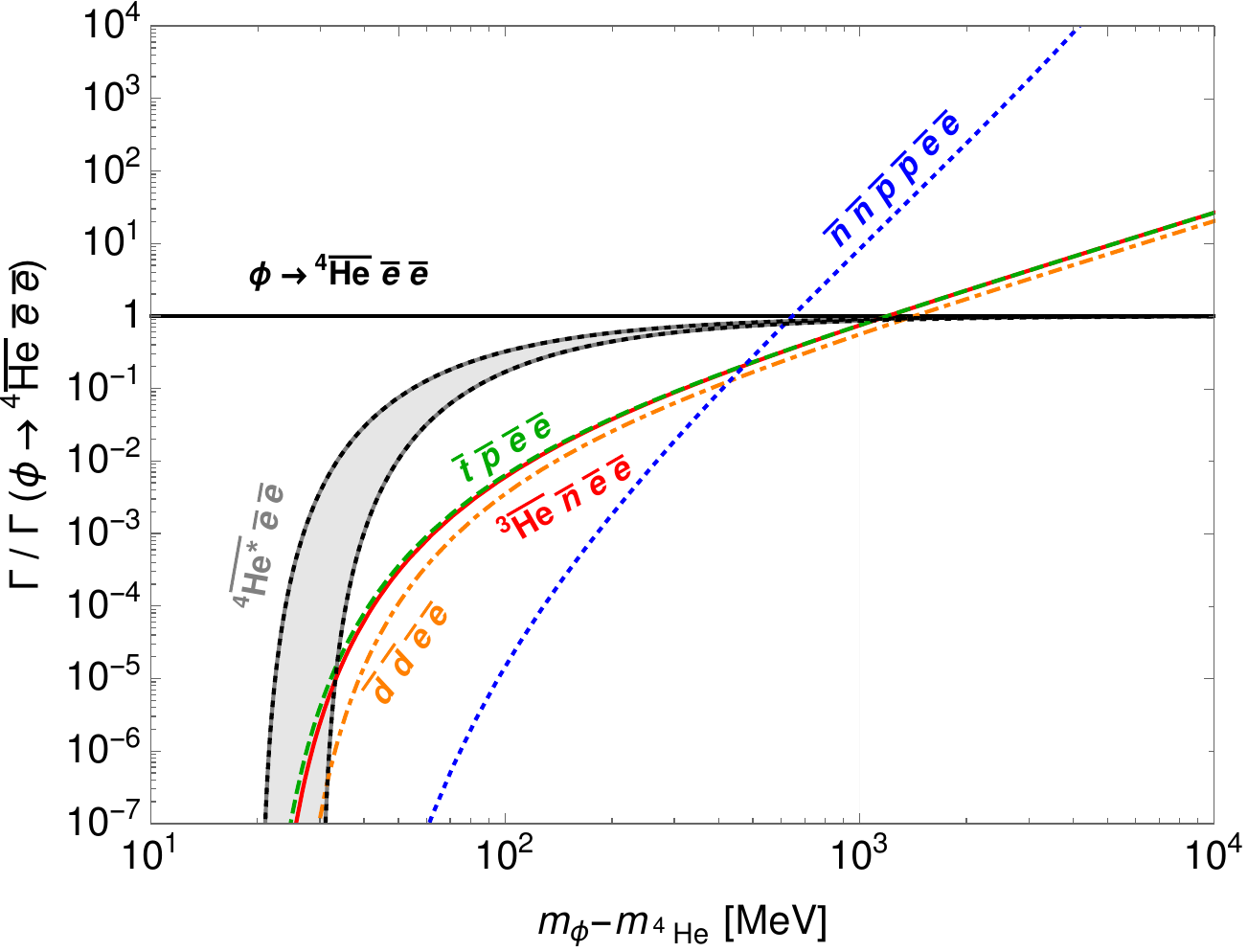}
\caption{Top: Lifetime of the decay $\phi \to \aHe{4}\, \anti{e} \,\anti{e}$ in Gyr 
for various effective-operator scales $\Lambda$.\\
Bottom: Other decay modes of $\phi$ relative to $\Gamma(\phi \to \aHe{4}\, \anti{e} 
\,\anti{e})$; in the gray-shaded region we expect $\He{4}^*$ excited states. 
We again denote $d\equiv {}^2\mathrm{H}^+$ and $t \equiv {}^3\mathrm{H}^+$.
All nuclear matrix elements are taken to be $\unit[100]{MeV}$ to the appropriate power.
The rates are calculated assuming $m_\phi-m_{\He{4}}\ll m_{\He{4}}$.
}
\label{fig:branchings}
\end{figure}

The nucleons are non-relativistic but the electrons relativistic, leading to the 
decay rate 
\begin{align}
\Gamma(\phi \to \aHe{4}\, \anti{e}\, \anti{e}) &\simeq \frac{ (m_\phi-m_{\He{4}})^5}
{3840 \pi^3\,m_\phi \,\Lambda^{12}} |\psi (0)|^6 \,.
\label{eq:decay_to_He4}
\end{align}  
We expect this expression to be valid for $m_\phi-m_{\He{4}}\ll m_{\He{4}}$.
With $|\psi (0)|^6\simeq (\unit[100]{MeV})^{9}$ this yields the lifetime 
\begin{align}
\tau &\simeq \unit[9\times 10^{7}]{Gyr} \left(\frac{\unit[10]{MeV}}{m_\phi-m_{\He{4}}}\right)^5
\left(\frac{\Lambda}{\unit[100]{GeV}}\right)^{12} ,
\end{align}
shown in Fig.~\ref{fig:branchings} (top). $\tau$~depends strongly on the 
mass splitting $m_\phi-m_{\He{4}}$ and the effective 
scale $\Lambda$, and can easily surpass the age of the universe, 
$t_\text{Universe}\simeq \unit[14]{Gyr}$.

For masses $m_\phi-m_{\He{4}} > \unit[20]{MeV}$, additional final states become 
energetically allowed, in particular the excited $\He{4}$ states of 
Tab.~\ref{tab:He4_states}. All of these excited states are unbound and 
unstable~\cite{Tilley:1992zz}, but we still expect some of them to contribute as 
resonances to the decay $\phi \to \aHe{4}^* \, \anti{e}\, \anti{e}$, followed by 
the fast decay of $\aHe{4}^*$ into $\aHe{4}$, $\aHe{3}$, $\anti{t}$, 
and $\anti{d}$, depending on the excited state.
This is then a production mode for $\aHe{3}$ and $\anti{d}$. 
Due to the resonant enhancement, the decay rates 
for $\phi \to \aHe{4}^* \, \anti{e}\, \anti{e}$ should be formally 
similar to Eq.~\eqref{eq:decay_to_He4}, naively replacing 
$m_{\He{4}}$ and $\psi(0)$ by the relevant excited mass and wave function. The latter is 
unknown to us, but we expect at least some of the excited states to have a 
similar matrix element as the ground state~\footnote{For the first excited state Refs.~\cite{Hiyama:2004nf,Horiuchi:2012ohc} calculate a nuclear radius around 3 times larger than that of the ground state, $r_{\He{4}}\simeq\unit[1.5]{fm}$, which we could interpret as a correspondingly smaller matrix element. Since no calculations are available for the other excited states we will not show the result here.}.
 As shown in Fig.~\ref{fig:branchings} (bottom), 
this allows some of the excited states to catch up to the 
$\phi \to \aHe{4} \, \anti{e}\, \anti{e}$ rate. While the exact details depend 
on the excited-state 
wave functions and branching ratios, we generically expect $\phi$ decays 
to produce similar amounts of $\aHe{4}$, $\aHe{3}$, and $\anti{d}$ for 
$m_\phi \sim m_{\He{4}} + \unit[100]{MeV}\simeq \unit[3.8]{GeV}$. 

Aside from the three-body decays into excited $\aHe{4}$ states there are 
\emph{four}-body decays into ground-state nuclei that should be considered, namely 
$\phi \to \anti{t}\, \anti{p} \, \anti{e} \,\anti{e}$, $\aHe{3}\,\anti{n} \, 
\anti{e} \,\anti{e}$, and $\anti{d} \,\anti{d}\, \anti{e} \,\anti{e}$ 
(Tab.~\ref{tab:thresholds}). 
All these four-body decays 
$\phi \to \bar A \, \bar B\,\anti{e} \,\anti{e}$ with non-relativistic 
$\bar A$ and $\bar B$ can be calculated using Fermi's golden rule,
\begin{align}
\Gamma (\phi \to \bar A \, \bar B\,\anti{e} \,\anti{e}) &\simeq 
\frac{|\psi(0)|^4}{270270 \pi^5 \Lambda^{12} m_\phi}
\left(\frac{m_A m_B}{m_A+m_B}\right)^{3/2}\nonumber\\
&\quad \times \left(m_\phi -m_A-m_B\right)^{13/2} ,
\end{align}
a factor $1/2!$ should be included for $A=B$. 
The nuclear matrix element or wave-function overlap $\psi(0)$ depends in principle on $A$ 
and $B$, but will here be approximated 
as before as $|\psi (0)|\sim (\unit[100]{MeV})^{3/2}$.
Under these approximations we find that the four-body decays never dominate the $\phi$ 
decay, as can be seen in Fig.~\ref{fig:branchings} (bottom).
Still, for $m_\phi-m_{\He{4}} \gtrsim \unit[1]{GeV}$ they could dominate over 
the $\aHe{4}$ production while still yielding similar numbers of 
$\aHe{3}$ and $\anti{d}$ overall, assuming our formulae remain valid in this 
region. More importantly, for $m_\phi-m_{\He{4}} < \unit[1]{GeV}$ we expect 
$\phi$ decays into excited $\aHe{4}$ states to dominate the production 
of $\aHe{3}$ and $\anti{d}$.

Finally, when $m_\phi$ is above the threshold for decay into free nucleons, 
$m_\phi >2m_\mathrm{n} + 2m_\mathrm{p}$, but still small enough to not resolve 
the underlying quark structure or heavier baryons such as $\Sigma$ and $\Delta$, 
$\phi$ will decay into free nucleons.
Using the matrix element from Eq.~\eqref{eq:msq} and \texttt{CalcHEP}~\cite{Belyaev:2012qa} 
to calculate the six-body phase space numerically we find the decay rate  described 
to excellent degree by
\begin{align}
\Gamma(\phi \to  \anti{n}\,\anti{n}\,\anti{p}\,\anti{p}\,\anti{e}\,\anti{e}) 
&\simeq \frac{3.3\times 10^{-19}}{m_\phi^6 \Lambda^{12}}\left(m_\phi 
+2m_\mathrm{n}+2m_\mathrm{p} \right)^{19/2}\nonumber\\
&\quad \times\left(m_\phi -2m_\mathrm{n}-2m_\mathrm{p} \right)^{19/2} .	
\end{align}
The decay to free nucleons dominates for large $m_\phi$ and  will eventually 
lead to constraints from antiproton-flux measurements that plague most DM models that aim to produce 
heavier antinuclei~\cite{Carlson:2014ssa,Cirelli:2014qia}.

The main conclusion from this analysis is that $\phi$ decays can produce 
almost exclusively $\aHe{4}$ for $0<m_\phi-m_{\He{4}} \ll \unit[100]{MeV}$, while for 
$m_\phi-m_{\He{4}} \sim \unit[100]{MeV}$ we expect 
similar amounts of $\aHe{4}$, $\aHe{3}$, and $\anti{d}$. For even 
larger $\phi$ masses, where our approximations become invalid, the $\aHe{4}$ fraction will decrease and the 
amount of $\anti{p}$ will become significant, eventually leading back 
to coalescence fractions for the antinuclei.

\subsection{\texorpdfstring{$\phi$}{phi} connection to  dark matter and UV completion}

In complete analogy to Sec.~\ref{sec:phiDM} we can imagine $\phi$ to be DM; the CMB bound $\tau > \unit[10^{25}]{s}$~\cite{Slatyer:2016qyl} now corresponds to $\Lambda >\unit[300]{GeV}$ for $m_\phi =\unit[3.8]{GeV}$, otherwise the same comments apply regarding potential acceleration mechanisms.
Exotic events of the form 
$\phi\,\mathrm{p}\to  \anti{p}\,\anti{n}\,\anti{n}\,\anti{e}\,\anti{e}$
in Super-Kamiokande are once again sufficiently suppressed.

Alternatively, in order to not rely on an exotic astrophysical $\aHe{4}$ acceleration mechanism we can imagine $\phi$ to be \emph{produced boosted} by a heavier DM particle $\chi$, either via annihilations or decays (e.g.~$\chi \to \phi\phi$ if 
$\chi$ carries baryon number $-8$ or $\chi \to \phi\phi\phi$ if $\chi$ has baryon number $-12$). Analogous to Sec.~\ref{sec:helium3subDM} this can easily give a large $\aHe{4}$ flux without accompanying antiproton or antideuteron flux, but requires sub-TeV $\Lambda$.

This brings us to the main issue of $\aHe{4}$ production from our model:
Eq.~\eqref{eq:lagrangian} has to come from a quark-level operator of dimension $\geq 22$ with scale $\Lambda_\mathrm{UV}\lesssim (\Lambda_\mathrm{QCD}/\Lambda)^{2/3}\, \Lambda$, which is at the GeV scale for the $\Lambda$ values of interest to us. $\Lambda_\mathrm{UV}$ is naively of order of the (colored) mediator masses  and thus most likely excluded by collider searches.
It is then more useful to study variations of our model along the lines of Sec.~\ref{sec:UV}, i.e.~considering decays $\phi_1\to\phi_2\,\aHe{4}\,\anti{e}\,\anti{e}$ near threshold with TeV-scale $\phi_j$ that can be resonantly enhanced by TeV-scale colored mediators.
While on the baroque side, models along these lines are so far the only explanation for $\aHe{4}$ events in AMS that do not rely on hidden antimatter regions~\cite{Poulin:2018wzu}.

\bibliographystyle{utcaps_mod}
\bibliography{BIB}

\end{document}